\documentclass[12pt,preprint]{aastex}

\def\kms{km s$^{-1}$}

\def\and{$\&$ }

\def\mjb{mJy beam$^{-1}$}
\def\asec{$^{\prime\prime}$}

\usepackage{wrapfig}

\begin{document}

\title{Ionized Gas Kinematics and Morphology in \\ Sgr B2 Main on 1000 AU Scales}

\author{C. G. De Pree}
\affil{ Department of Physics and Astronomy, Agnes Scott College}
\affil{141 E. College Ave., Decatur, GA 30030}
\email{cdepree@agnesscott.edu}
\author{D. J. Wilner}
\affil{Harvard Center for Astrophysics}
\affil{Cambridge, MA}
\email{dwilner@cfa.harvard.edu}
\author{W. M. Goss}
\affil{National Radio Astronomy Observatory}
\affil{Socorro, NM}
\email{mgoss@nrao.edu}

\begin{abstract}
We have imaged the Sgr B2 Main region with the Very Large Array in the BnA configuration ($\theta_{beam}$ = 0\farcs13) in both the H52$\alpha$ (45.453 GHz) radio recombination line (RRL) and 7 mm continuum emission. At a distance of 8500 pc, this spatial resolution corresponds to a physical scale of 0.005 pc ($\sim$1100 AU). The current observations detect H52$\alpha$ emission in 12 individual ultracompact (UC) and hypercompact (HC) HII regions. Two of the sources with detected H52 $\alpha$ emission have broad ($\Delta$V$_{FWHM}\sim$50 \kms) recombination lines,  and two of the sources show lines with peaks at more than one velocity. We use line parameters from the H52$\alpha$ lines and our previous H66$\alpha$ line observations to determine the relative contribution of thermal, pressure and kinematic broadening, and electron density. These new observations suggest that pressure broadening can account for the broad lines in some of the sources, but that gas motions (e.g. turbulence, accretion or outflow) contribute significantly to the broad lines in at least one of the sources (Sgr B2 F3).
\end{abstract}

\section{INTRODUCTION}
The Sgr B2 star forming region, located near the Galactic center is one of the most luminous in the Galaxy, and is associated with a 10$^6$ M$_{\odot}$ giant molecular cloud (GMC). The region is highly extincted at optical and infrared wavelengths, but has been extensively studied at radio wavelengths (Qin et al. 2011, De Pree et al. 1998, Gaume et al. 1995). Gaume et al. (1995) published the first high resolution ($\theta_{beam}$=0\farcs25, $\sim$2000 AU) radio images of the Sgr B2 Main, South and North star forming regions. These original 1.3 cm Very Large Array (VLA) continuum images were followed by H66$\alpha$ (1.3 cm) radio recombination line (RRL) observations at the same resolution, lower resolution ($\theta_{beam}$ = 2\farcs5) H52$\alpha$ (7 mm) RRL observations (De Pree et al. 1996) and high resolution ($\theta_{beam}$=0\farcs065, $\sim$600 AU) 7 mm continuum observations (De Pree et al. 1998).  These final 7 mm continuum observations revealed complex morphologies for a number of the sources first imaged at 1.3 cm. Our previous H52$\alpha$ line observations (De Pree et al. 1996) had insufficient spatial resolution and sensitivity to determine RRL parameters for the individual 7 mm continuum sources discussed in De Pree et al. (1998). 

Galactic UC HII regions typically have high frequency recombination line widths of less than 25 \kms~ (Osterbrock 1989). For example, the thermal width, assuming a constant temperature inside an HII region of 8000 K, is 19.1 \kms~ (Keto et al. 2008). Jaffe \& Martin-Pintado (1999) found in a survey of Galactic UC HII regions that a substantial fraction of the surveyed sources ($\sim$30 \%) had both radio recombination lines that were significantly broader than the typical thermal profiles ($\Delta V_{FWHM} > $ 50 \kms), and rising spectral indices ($\alpha$ $>$ 0.4, where S$_{\nu}$ = $\nu^{\alpha}$). They designated such objects broad recombination line objects, or BRLOs. De Pree et al. (2004) found a similar fraction of BRLOs in their 7 mm recombination line and continuum study of W49A. The exact physical process that accounts for the presence of BRLOs is unclear, though the combination of kinematically broadened lines and rising spectral indices is consistent with ionized outflow, perhaps from a circumstellar disk, several examples of which have been detected, e.g. K3-50A (De Pree et al. 1994). 

Recent work by Keto et al. (2008) suggests that some of the broad lines in Jaffe \& Martin-Pintado (1999) are the result of pressure broadening, and that pressure broadening can have a significant effect, even at frequencies as high as 45 GHz. However, Keto et al. (2008) conclude that some sources certainly have broad lines related to gas motions, and that multifrequency RRL observations can be used to determine the relative contributions of thermal, pressure and kinematic broadening processes. Some of the broad lines in these sources may also be explained by a proposal that UC HII region lifetimes and morphologies may be the result of source ``flickering'' (e.g. Peters et al. 2010). The flickering is modeled as a result of the trapping of ionized accretion flows (Galv\'an-Madrid et al. 2010). Whether the broadening is due to high densities (pressure broadening) or gas motions, this subset of UC HII regions represents a distinct phase in the early lifetime of a forming massive star. The current 7 mm RRL observations indicate that in Sgr B2 Main, very broad recombination lines do persist in a number of sources on small scales ($\sim$1000 AU), and at high frequencies ($\sim$45 GHz).

In this paper, we discuss the H52$\alpha$ line parameters from new RRL observations of 12 individual  sources in Sgr B2 Main imaged at the same resolution as the 7 mm continuum (De Pree et al. 1998), and determine the contributions to the detected line widths that arise from thermal broadening, pressure broadening, and unresolved gas motions. We also determine the electron density (n$_e$) from the line widths of RRLs measured at two frequencies, and spectral indices between 7 mm and 1.3 cm. The spectral indices for the sources in this study are derived from the 1.3 cm flux density data from Gaume et al. (1995) and 7 mm flux density data from the current work. 

\section{OBSERVATIONS and RESULTS}
Observations of the Sgr B2 Main region were made with the VLA\footnote{The National Radio Astronomy Observatory is a facility of the National Science Foundation operated under cooperative agreement by Associated Universities, Inc.} in the BnA configuration on 4 October 2003. The flux calibrator was B1328+307, which was used to determine the  flux density of the phase calibrator, B17425-28593. We calculated that the flux density for our phase calibrator was 1.0$\pm$0.16 Jy. At the time of the observations, there were 27 antennas equipped with 7 mm receivers, with a total of 2 hours 25 minutes on source. Observations were made with fast switching, with observations cycling rapidly between the source (60 s) and the phase calibrator B17425-28593 (20 s). The VLA correlator setup provided a 25 MHz bandwidth in 31 channels (no Hanning smoothing), and a channel separation of 781.25 kHz. The bandpass was centered on the H52$\alpha$ line with a rest frequency of 45.453 MHz (shifted to +65 \kms). The resulting spectra have  a velocity resolution of 6.2 \kms~and a velocity coverage of 160 \kms. In the calibration process, 3-4 end channels were lost due to bandpass calibration effects, providing an effective velocity coverage of about 125 \kms.

\begin{figure}[t]
\begin{center}
    \includegraphics[width=0.85\textwidth]{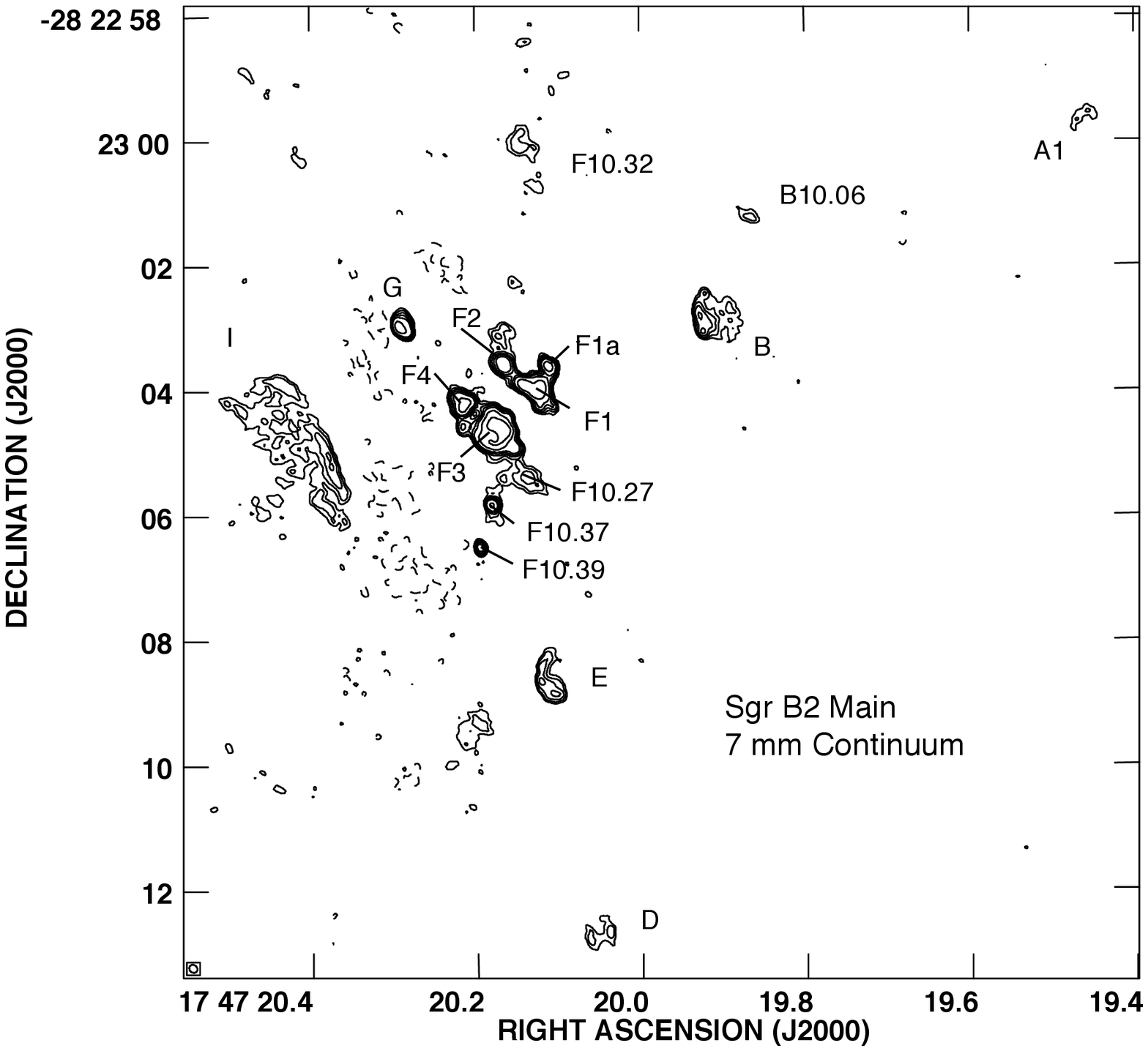}
  \end{center}
\caption{The 7 mm uniformly weighted continuum image of Sgr B2 Main. The first contour level is 4$\sigma$ level (4.1 mJy/beam). Successive contours are at 1.4, 2, 2.8, 4, 8, and 16 times the 5$\sigma$ level. Restoring beam is $\theta_{beam}$= 0.14\arcsec $\times$ 0.12\arcsec, P.A. = 46$^{o}$). Peak flux density in field is 0.11 mJy/beam. }
\end{figure}

\begin{figure}[t]
\begin{center}
    \includegraphics[width=0.75\textwidth]{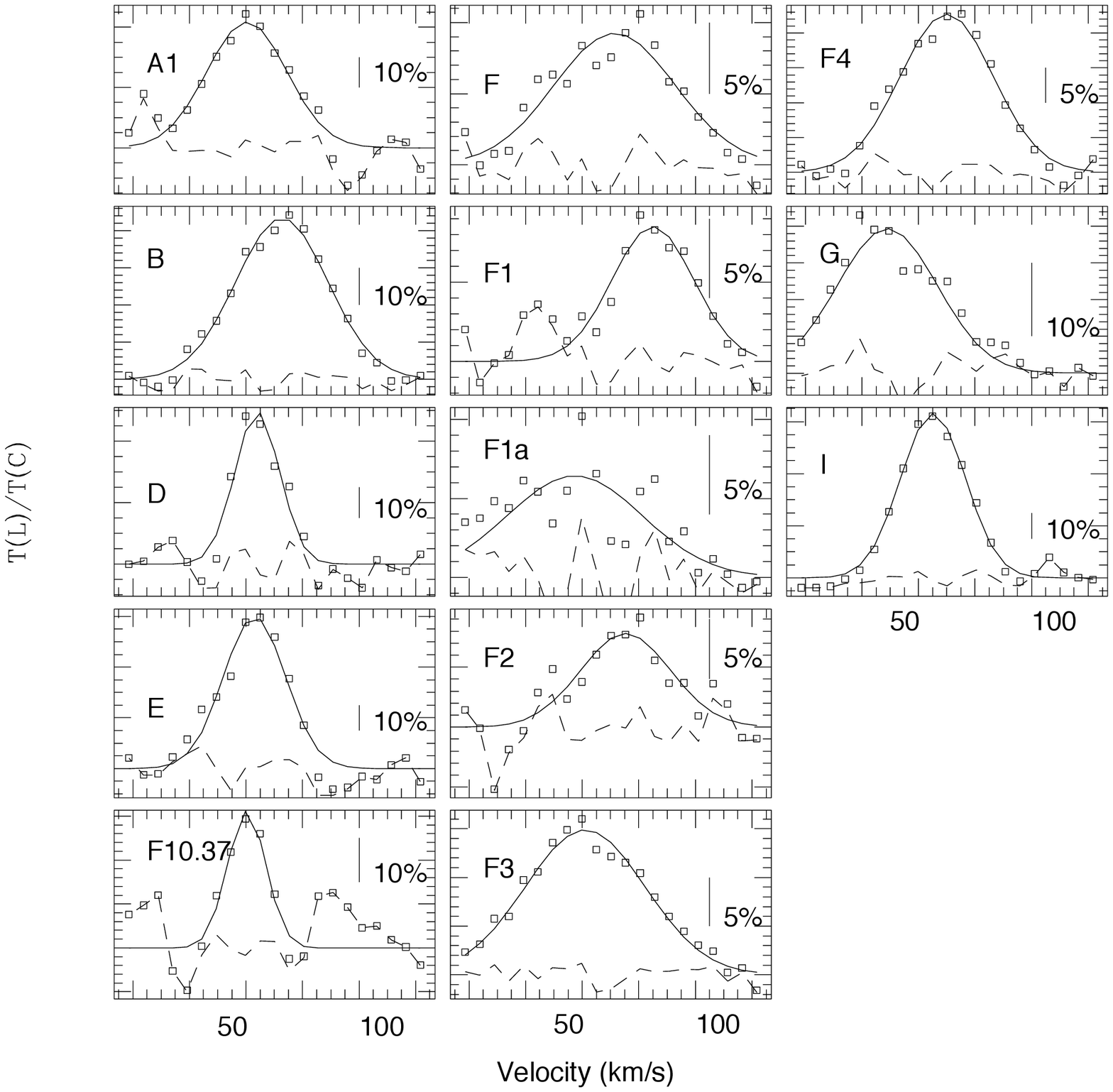}
  \end{center}
\caption{H52$\alpha$ (45.4537215 GHz) line emission from the continuum sources in Sgr B2 Main. Sources are labelled with the continuum source name. Plot for each source shows data (square), single Gaussian fit (solid line) and residual (dotted line). Bars in each plot indicate 10\% (line to continuum ratio) except for the F sources (5\% as indicated). Line parameters of the H52$\alpha$ line are listed in Table 1, along with the H66$\alpha$ line parameters from De Pree et al. (1996)}
\end{figure}

Data were reduced using standard techniques with AIPS data reduction software, 
and the continuum image was self calibrated. Calibrations from the continuum 
data were applied to the line data. Data were imaged with the AIPS task IMAGR, 
and the restoring beam was 0.14\asec x 0.12\asec, PA = 46$^o$. The rms noise in the 
continuum data was 1.0 \mjb~and the rms noise in the line data was 6.6 \mjb.

Line parameters were fitted in the Groningen Image Processing System (GIPSY), using the task PROFIL. The data were averaged over each source and fitted with a Gaussian. From this Gaussian, the line velocity (V$_{LSR }$), full width at half maximum ($\Delta$V$_{FWHM}$), and line to continuum ratio (${T_L}\over{T_C}$) were determined. Several of the sources had multiple peaks. Due to the low spectral resolution of the data, these were fit as best as possible with a single Gaussian. Keto et al. (2008) discuss the merits of fitting Voigt versus Gaussian profiles to RRLs, and conclude that Gaussian profiles provide the best fit in the peak of the line, without overly weighting the low signal to noise data in the line wings. Table 1 summarizes the fitting results for the current  H52$\alpha$ data, and the H66$\alpha$ line data (1996). Column (1) gives the source name, columns (2)-(3)  gives the line width to half-power, columns (4)-(5) give the line velocity (\kms), and columns (6)-(7) give the line to continuum ratio in each of the lines.

Three of the sources detected in the continuum (Sgr B2 B10.06, F10.27, and F10.32) do not have detected line emission in the H52$\alpha$ line. Non-detections for these UC HII regions may be the result of RRLs that are too broad for the bandwidth of these observations. Very broad lines have been detected in some regions, with H66$\alpha$ line widths, for example, of $\sim$180 \kms~(Keto et al. 2008). Several UC sources in the F region that had very broad lines at 1.3.cm (e.g. F2 and F4) have 7 mm recombination lines that are broader than thermal, but not as broad as the line detected at 1.3 cm, suggesting that in some sources, the broad lines detected at 1.3 cm may be due to pressure broadening.

\subsection{Spectral Indices}
As the result of  high densities (n$_e$ $>$10$^5$ cm$^{-3}$) in the inner regions of sources with density gradients maintained by the gravity of forming massive stars, the density gradient can produce a rising spectral index. Spectral index measurements are sensitively dependent on the uv-coverage of the observations, with large arrays being less sensitive to larger structures. The 1.3 cm continuum (Gaume et al. 1995) and 7 mm continuum (De Pree et al. 1998) used for these measurements have similar spatial resolutions, but are not well-matched in uv-coverage. The 1.3 cm data (Gaume et al. 1995) are the result of observations in 3 hybrid arrays (DnC, CnB and BnA), while the 7 mm continuum and line data were made with observations in the BnA configuration alone. As a result, the higher frequency 7 mm observations are less sensitive to large scale structure, meaning that flux density measurements at 45.453 MHz for extended sources will be lower limits, and the spectral index ($\alpha$) values for these sources are also lower limits.

We have determined a lower limit to the spectral index ($\alpha$) for the 15 detected continuum sources. These lower limits are listed in the final column of Table 1. Of these 15 continuum sources, we have determined that 4 of the sources $-$ all of which have small linear sizes, and are thus less effected by reduced sensitivity to large scale structures $-$ are consistent with rising continuum spectral indices between 1.3 cm and 7 mm. Sources Sgr B2 F3, F10.27, F10.32, and F10.37 have lower limits to their spectral index values (S$_{\nu}$ = $\nu^{\alpha}$) of 0.3, 1.4, 1.0 and 0.8 respectively. Of these 4 rising spectrum sources, only Sgr B2 F3 has a detected broad radio recombination line emission ($\Delta$V$_{FWHM}>$ 50 \kms). 

\subsection{Pressure Broadening and Electron Densities}
Broad RRLs are the result of a combination of thermal broadening, pressure broadening and kinematics (e.g. outflows and/or accretion). Keto et al. (2008) have examined the relative contributions of these three factors in a small number of bright, well-studied UC and HC HII regions. Based on their analysis of five regions, they conclude that a combination of gas motions and pressure broadening (resulting from the very high electron densities in the cores of some of these sources) are responsible for the broad lines in these sources. Radio recombination lines may experience pressure 
broadening in the H66$\alpha$ line if there is sufficient density ($>$10$^7$ cm$^{-3}$). However,  broad lines have been detected even in the H30$\alpha$ line (e.g. Keto et al. 2008), a frequency at which pressure broadening would be negligible for densities less than 10$^{8}$ cm$^{-3}$, indicating that in these sources, the broad lines are primarily kinematic.

Using their equation (3), we calculate that the contribution of pressure broadening at 45.453 GHz is negligible as long as the electron densities are less than 4.5 x 10$^6$ cm$^{-3}$. We have performed a similar analysis using the two observed RRLs in the Sgr B2 Main region. We have used fits to the H52$\alpha$ and the H66$\alpha$ lines from Table 1 in order to determine the contributions for each source from thermal broadening, dynamical broadening and pressure broadening. In our analysis, we use equations (1), (2) and (3) from Keto et al. (2008), and are able to determine the electron density from the H66$\alpha$ line. For each source, the derived width due to pressure broadening ($\Delta\nu_{L}$) and gas motions ($\Delta\nu_{d}$) are both listed in Table 1 in column 8 and 9 respectively. We also list the electron density (n$_e$) derived from the change in line width between the H52$\alpha$ and the H66$\alpha$ lines in column 10. For these calculations, we have assumed a thermal width (T=8000 K) of 19.1 \kms.

\section{DISCUSSION}
Twelve of the fifteen continuum sources in Sgr B2 Main have detected RRL emission in the H52$\alpha$ line. Three (3) of the sources with detected line emission (Sgr B2 A1, B and G) do not appear to have significant pressure broadening of the H66$\alpha$ line, both from the fact that the lines to not significantly narrow from the H66$\alpha$ to the H53$\alpha$ line, and from the fact that analysis shows that pressure broadening contributes negligibly to the width of the H52$\alpha$ RRL. Morphologically, these three sources are also relatively extended. 

Three (3) of the detected sources with RRL emission (Sgr B2 F1, F2, and F4) have lines that narrow significantly between 1.3 cm and 7 mm. An analysis of the change in line width from H66$\alpha$ to H52$\alpha$ indicates that pressure broadening is a significant contributor to the width of the RRL in these three sources at the frequency of the H66$\alpha$ line (see Table 1).  

Finally, four (4) sources (Sgr B2 D, E, F3, and I) are marginally effected by pressure broadening in the H66$\alpha$ line (see Table 1), with thermal and pressure broadening contributing at approximately the same level. Of these four sources, only one (Sgr B2 F3) is detected to have a very broad RRL ($\Delta$ V $>$ 50 \kms) at 7 mm. The other three sources have typical RRL line widths at 7 mm, indicating that their lines at 45.453 GHz are not pressure broadened at 7 mm. 

Of the sources detected in Sgr B2 Main only F3 has both a rising continuum spectral index, and a broad radio recombination line that does not appear to be significantly effected by pressure broadening. Thus, Sgr B2 F3 could be categorized as a Broad Recombination Line Object (BRLO), as defined by Jaffe \& Martin-Pintado (1999). Sgr B2 F3 also has the largest derived width due to gas motions ($\nu_d$=46 \kms). The morphology of source F3 is shell-like.

Only one other source (Sgr B2 F1a) has an RRL that is broad ($\Delta V_{FWHM} > $ 50 \kms) at 7 mm, but there is no H66$\alpha$ RRL detection for comparison. Further observations with much broader bandwidth and better spectral resolution, as provided by the recently commissioned EVLA correlator, are required to understand the nature of the radio recombination line emission in the remainder of these sources.

\acknowledgements{This research was supported by a grant from the National Science Foundation (AST-0206103). Some of the work on this paper was completed while C. De Pree was a visiting scientist on sabbatical at Emory University Department of Physics, and he gratefully acknowledges the Department.}

\begin{deluxetable}{lccccccccccccc}
\tabletypesize{\tiny}

\tablenum{1}
\tablecolumns{14}
\tablecaption{Radio Recombination Line Parameters}

\tablehead{
Source & \multicolumn{2}{c}{$\Delta$V$_{FWHM}$} & &  \multicolumn{2}{c}{V$_{LSR}$ (km/s)} & &\multicolumn{2}{c}{T$_{L}$/T$_{C}$ (\%)} & &  \multicolumn{3}{c}{Derived Parameters}\\
\cline{2-3} \cline{5-6}  \cline{8-9} \cline{11-14}\\
& \colhead{H52$\alpha$} & \colhead{H66$\alpha$} & &  \colhead{H52$\alpha$} & \colhead{H66$\alpha$} & &  \colhead{H52$\alpha$} &  \colhead{H66$\alpha$} & & \colhead{$\Delta\nu_L$} & \colhead{$\Delta\nu_d$}& \colhead{$n_e$}&\colhead{$\alpha\tablenotemark{a}$}\\
& & & & & & & & & &\colhead{(km/s)} & (km/s) &(10$^6$ cm$^{-3}$) & (S$_{\nu}\propto\nu^{\alpha}$)}

\startdata
A1 & 34$\pm$4 & 36.4$\pm$1.2&&60$\pm$2&64.5$\pm$0.4&&42$\pm$4&26$\pm$1&&4.4&28&0.2 &-4.8 \\
B&38$\pm$1&40.7$\pm$1.3&&72$\pm$1&71.1$\pm$0.5&&43$\pm$1&16$\pm$1&&4.9&33&0.22&-0.5\\
B10.06&-&35.7$\pm$7.2&&-&46.3$\pm$2.6&&$<$73&34$\pm$10&&-&-&-&-0.1\\
D&19$\pm$2&34.1$\pm$1.3&&64$\pm$1&63.9$\pm$0.5&&50$\pm$4&21$\pm$1&&23&-&1.0&-1.0\\
\\
E&26$\pm$2&34.8$\pm$1.0&&63$\pm$1&62.3$\pm$0.4&&60$\pm$4&21$\pm$1&&14.8&18&0.66&-1.4\\
F1&34$\pm$4&61$\pm$12&&85$\pm$2&79$\pm$5&&6$\pm$1&1.3$\pm$0.31&&41.1&28&1.8&-0.6\\
F10.303 (F1a)&57$\pm$10&-&&57$\pm$5&-&&6$\pm$1&-&&-&-&-&-0.6\\
F2&38$\pm$5&78$\pm$10&&75$\pm$2&68$\pm$4&&8$\pm$1&5.0$\pm$1.01&&58.5&33&2.6&-0.9\\
\\
F3&50$\pm$3&63$\pm$5&&61$\pm$1&68$\pm$2&&15$\pm$1&3.0$\pm$0.3&&22.3&46&1.0&0.3\\
F4&39$\pm$1&59$\pm$6&&70$\pm$1&72$\pm$2&&23$\pm$1&6.0$\pm$0.7&&32.2&34&1.4&-0.3\\
F10.27&-&-&&-&-&&$<$20&-&&-&-&-&1.4\\
F10.32&-&-&&-&-&&$<$24&-&&-&-&-&1.0\\
\\
F10.37&16$\pm$3&-&&60$\pm$1&-&&31$\pm$5&-&&-&-&-&0.8\\
G&42$\pm$4&44.3$\pm$5.1&&49$\pm$1&46.7$\pm$2.1&&20$\pm$2&7$\pm$1&&4.2&37&0.19&-0.3\\
I&27$\pm$1&36.6$\pm$0.8&&65$\pm$1&60.8$\pm$0.3&&63$\pm$2&19$\pm$1&&16.1&19&0.72&-2.5\\

\enddata
\tablenotetext{a}{Spectral index values listed are lower limits, calculated between 22.4 GHz and 45.5 GHz. These are lower limits due to potential missing flux at the higher frequency.}

\end{deluxetable}

\end{document}